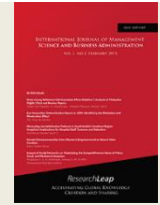

# The Effects of Using Business Intelligence Systems on an Excellence Management and Decision-Making Process by Start-Up Companies: A Case Study

[1] Otmane Azeroual, [2] Horst Theel

[1] German Center for Higher Education Research and Science Studies (DZHW), Berlin, Germany
[2] University of Applied Science HTW Berlin, Department of Computer Science, Communication and Business, Berlin, Germany

**Abstract:** The rapid increase in data volumes in companies has meant that momentous and comprehensive information gathering is barely possible by manual means. Business intelligence solutions can help here. They provide tools with appropriate technologies to assist with the collection, integration, storage, editing, and analysis of existing data. While almost only large companies were interested in this topic a few years ago, it has meanwhile also become necessary for start-up companies, and so the market for business intelligence has been growing for years. This article focuses on the general potentials of using BI in start-ups. First, will be examined which providers of BI solutions are suitable for start-ups and what opportunities exist for implementing BI systems in start-ups. Then it will be shown to what extent BI has prevailed in start-ups, in which areas the techniques of BI are used in start-ups and what purpose BI has in start-ups. Finally, the success factors for BI projects in start-ups are considered.

## 1. Introduction

With increasing globalization of markets, fierce competition, increasing the speed with changes in market conditions and customer needs, all market participants and companies face new challenges. In the long run, companies will be able to assert themselves, who can adapt to these conditions, who can respond flexibly and quickly to changes while at the same time keeping their costs under control. For this purpose, however, an exact knowledge of the current corporate and market situation is indispensable. To ensure this and to provide management with the information needed in their planning and decision-making, sophisticated information and communication systems are used. Since the 1960s, various approaches have been developed for such systems, which have become known under many different names such as Management Information Systems (MIS), Decision Support Systems (DSS) or Executive Information Systems (EIS). Today, the term Business Intelligence (BI) has become established both in practice and in research. BI describes approaches such as collecting, storing, processing, analyzing and presenting company data.

In recent years, Business Intelligence has become one of the top topics in the German and international IT market. In this regard, the importance of companies has increased significantly. By using BI systems, companies are supported in making their business-critical data and processes transparent and intelligent. Also, employees will be able to make better decisions, achieve the required results faster, and continuously develop them. Another advantage of BI systems is that companies can make their customer and supplier relationships even more profitable, reduce costs, minimize risks and increase added value. Without the use of BI systems, enormous amounts of data are available, but then they spread confusion and ultimately complicate business.

A few years ago, almost only large companies and corporations showed interest, but now this topic has become increasingly attractive for start-up companies. Because there they discovered what potential could be tapped with Business Intelligence. One reason for this is that BI software has become increasingly cheaper and more affordable for many start-ups. For example, the BI market is flooded by software vendors (such as SAP, Oracle, IBM, SAS, Microsoft, and open source vendors) that are specific to start-ups. On the other hand, increasing competitive pressure and the requirement to be able to rely on reliable information quickly and at all times ensure strong demand.

Against this background, the aim of this present paper is to demonstrate the use of business intelligence in start-ups and to give an overview of the providers of business intelligence solutions that are suitable for start-up companies.

## 2. Business Intelligence

The term Business Intelligence (BI) was introduced by Gartner Group analyst Howard Dresner in the middle of 1990s and defined as a collective term for concepts and methods that support decision-making through information analysis, delivery, and processing. BI has become widespread in business practice and science and is widely used. However, there is still disagreement in understanding the term. This uncertainty leads to an undefined variety of definitions. A precise demarcation proves to be difficult since each selected definition remains vulnerable. In 1996, Business Intelligence was defined as follows: „Data analysis, reporting, and query tools can help business users wade through a sea of data to synthesize valuable information from it – today these tools collectively fall into a category called Business Intelligence" (Anandarajan, Srinivasan, and Anandarajan, 2004).

Due to the different understanding of business intelligence, different architectures for business intelligence systems are presented in the literature. About the broad knowing of the term used in the present work, above all, various logical processes are given in the references, which forms the basis for a BI architecture. These processes are assigned to the individual concepts and techniques that are summarized in the term Business Intelligence.

In this paper, the following processes will be distinguished in BI architecture:
1. Data Collection
2. Data Integration
3. Data Storages
4. Data Processing
5. Data Presentation

The data collection includes the operational systems that provide the required data for the Business Intelligence system. In particular, a distinction must be made here between internal and external systems as sources. Through data integration, the required data is transferred from pre-systems, processed and condensed, which is referred to as ETL process. The purpose of the ETL process is to ensure that the processed data can be stored persistently in the data storage or maintenance. The data storage can be realized in different architectural variants. Here Data Warehouse and Data Marts are used. In data processing or data analysis, all concepts and tools that are primarily concerned with the evaluation and analysis of the data are assigned to this process. This level is therefore assigned to analytical applications, which evaluate the data stored in the data storage process according to predetermined criteria. This process also includes components that enable online analytical processing (OLAP) and data mining components that are used to detect data patterns. In the data presentation is the target group specific preparation and presentation of the analysis results for the user. For this purpose, different concepts are used, such as OLAP clients for the implementation of ad-hoc inquiries or prefabricated target-group-specific reports. This level can also be assigned dashboards or management cockpit, planning and balanced scorecards, which are becoming increasingly important.

The following figure gives an overview of the individual processes and shows which components belong to which process step.

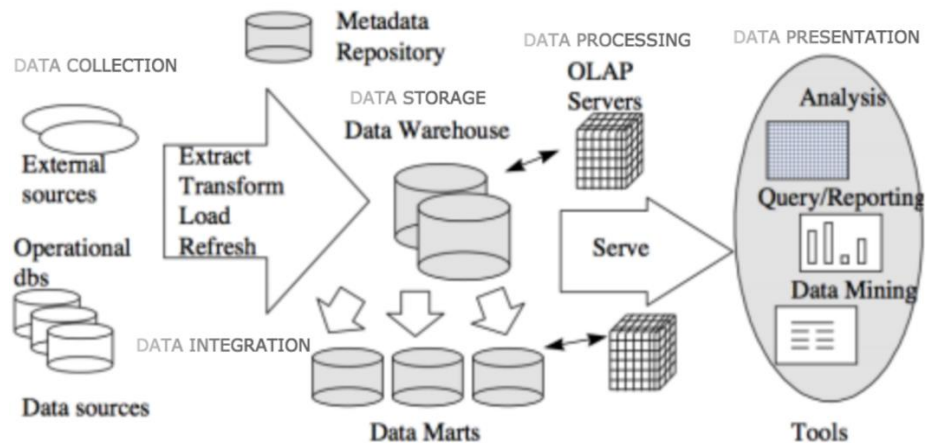

**Figure 1:** Business Intelligence Process (Wang, J., Chen, T. and Chiu, S. (2005))

# 3. Objectives of Business Intelligence

In business, it is important to have sufficient, high-quality information and key performance indicators (KPIs) as the basis for decisions. The biggest problem, however, is primarily the merging of information. Many documents, spreadsheets or databases hide inside companies and contain a lot of relevant information that is very difficult to merge and provide meaningful information. So, an overall holistic view is to be created, which is available as a basis for decisions. Therefore, it is possible, for example, optimize business processes, minimize risks, reduce costs, increase added value. Also, the use of BI can become a real competitive advantage, since the necessary information is practically available at the click of a mouse. Using Business Intelligence has three main objectives.

Improvement of decision basis: The decisions to be taken are usually made by available information. Thus, it is obvious that with better preparation of information as well as with the consideration of a multiplicity of information the basis of the decision is substantially improved. Information, that is, facts about certain things, is present in large scale in today's businesses. The use of BI makes it possible to translate them into a format that gives them an information advantage in their daily work.

Increasing the transparency of corporate actions: With the help of BI, the employee should be enabled to take responsibility for his area through facts and operating numbers and to be able to understand company decisions. BI provides the opportunity to enhance that visibility and empowers employees to see the impact of their area directly in the context of the enterprise as a whole.

Demonstrate the relationships between solitary information: Due to the complexity of business processes, decisions in many areas of companies have far-reaching consequences. The BI solutions' aim to link data from different sources and to recognize relationships that cannot be obtained from the individually considered sources of information. This is the knowledge component of BI, which should be carried out by absolute professionals in the relevant field, as generating such information increases the transparency of the entire process within companies.

# 4. Requirements of Business Intelligence

Almost every business can benefit from the use of business intelligence, but there are not always the right conditions to successfully implement business intelligence. It's not just about the technical prerequisites to be able to access the relevant databases, but also questions of corporate culture and the way in which business intelligence is approached. Essentially, the three requirements for BI can be represented as follows:

Willingness to make things in question: BI offers the opportunity to challenge things that have been in the business for a long time. It is possible to analyze changed structures, to create new information combinations and to simply look at changes on suspicion. If the basic structure of the information is well-structured, regarding both the technical conditions and the organization of the data, this offers a wide field of confirmation.

Willingness to think unconventionally: Business Intelligence creates entirely new insights. These must be explicitly allowed to take into account things that might at first glance be considered nonsensical or unrelated.

Management Attention: From the first two points, realize that BI projects have to take one or the other way to be successful. However, this also means that the management also accepts this. It is very administrable for BI projects if they are provided with the appropriate support from the responsible management, this significantly increases the acceptance and willingness to participate in such processes.

# 5. Business Intelligence Providers for Start-ups

Although business intelligence solutions are seen as major investments for large companies, more and more providers have been pushing for start-ups for several years. In the following, the solutions of the larger providers are presented briefly:

### SAP

The German software provider SAP has been a market leader in Germany for years and internationally at BI-Software. "SAP BusinessObjects Edge" offers a comprehensive and powerful business intelligence solution for start-ups. The product includes features for different BI needs: flexible ad-hoc queries and analytics, dashboards and visualization, data integration, and pre-configured data mart solutions. SAP BusinessObjects Edge Standard, SAP BusinessObjects Edge with Data Integration and SAP BusinessObjects Edge with Data Management are available.

### Oracle

The "Business Intelligence Standard Edition one" of Oracle BI system is geared to the needs of start-up companies. As a complete solution, this is designed for five to fifty users. Comprehensive features such as interactive dashboards, ad-hoc analytics, proactive intelligence and insight, and advanced reporting and publishing mobile and predictive analytics are included in the solution.

### IBM

IBM Cognos products provide business intelligence solutions for start-ups. The tools for start-ups include the functions Reporting, Analysis, Planning, Budgeting, and Forecasting as well as Dashboards and Balanced Scorecards.

### SAS

With the SAS Business Intelligence - Edition M product, SAS has launched a bespoke BI solution for start-ups. The solution has a modular structure and can, therefore, be expanded as requirements grow. The Edition M consists of a data integration component, an add-in for integration with Microsoft Office products, a dynamic desktop interface, and a Web Report Studio.

### Microsoft

With "Microsoft Reporting and Analysis Services" Microsoft provides a Business Intelligence solution for start-ups. The solution offers data presentation and reporting, comprehensive analysis functions, fast access to large volumes of data and extensive functionality in the Microsoft standard software.

### Open source vendors

In addition to big players such as SAP, Oracle, and IBM, some open source providers such as Jaspersoft, Pentaho, Jedox, and SpagoBI have established themselves on the market in recent years, offering solutions for start-ups. These four vendors use a similar business model that builds on commercial open source software. It includes a Free Community version and a Paid Enterprise version with vendor support and special features. Nevertheless, the four open source subjects differ from each other:

### Jaspersoft

Jaspersoft is one of the largest providers of integrated BI solutions for start-ups. Various modules are offered for the individual subtasks, which can be connected to each other via interfaces without much effort. The list of individual modules is provided by Jaspersoft ETL, JasperReports Server (Report Server) with JasperReports Library and Jaspersoft OLAP (Mondrian Custom Solution), and Jaspersoft Studio (Report Editor). Jaspersoft relies on third-party vendors in the ETL and OLAP areas, but the modules are adapted to their product line. All submodules can also be used separately. Jaspersoft focuses on reporting and distribution. The report server is at the heart of the solution.

### Pentaho

Pentaho is one of the market leaders in open source BI for start-ups and has a high profile. Like Jaspersoft, Pentaho relies on customized, partially existing Open Source projects acquired by Pentaho. The focus is on data integration and reports automation. The product portfolio consists of Business Analytics Platform, Data Integration, Report Designer, Aggregation Designer & Schema Workbench and Metadata Editor.

### Jedox

Jedox offers a complete BI suite that includes everything from ETL to OLAP to dashboards and reports. The hobby was and still is the powerful OLAP module Jedox. By integrating into Excel, Jedox offers a lot of added value for many start-ups, since a wide variety of data is often already maintained in Excel spreadsheets. Jedox integrates with Excel and replaces the pivot function, but offers many additional features such as target / actual deviations. The greatest advantage of integrating with Excel is above all in the "familiar" environment, because office applications are widely used in start-ups. As a result, the training effort here is relatively low overall.

### SpagoBI

As the only provider of integrated BI solutions for start-ups, SpagoBI offers all modules only in an open source version. There is no Enterprise version of SpagoBI. The business model is to provide services in the field of module setup and customization. Essentially, SpagoBI offers the advantage that all tendered software solutions can be managed via a central platform via the browser. So can a solution that has been developed with SpagoBI. SpagoBI is a small BI suite made up of several open source BI tools packed together, e.g., ETL, OLAP, Data Mining, Reporting (including Talent, Mondrian, Weka, BIRT and Jasper-Reports Library). Also, SpagoBI does not offer further modules. They are mainly used to optimize the presentation of BI results on mobile devices or to support the connection of geographical data with business-relevant data.

In addition to the five large providers SAP, Oracle, IBM, SAS and Microsoft, start-up companies like to buy software providers from Open Source, such as Jaspersoft, Pentaho, Jedox Palo and SpagoBI, because they negotiate at eye level, have advantages in terms of local presence and support or know-how for certain tasks or industries.

In summary, it can be said that the providers of open source BI solutions for start-ups are well represented. In addition to a free version, they also offer a commercial version of their software. However, their business version can often be purchased cheaper than that of the significant providers. The difference between the free version and the paid version is in most cases in advanced features and professional support.

## 6. Use of Business Intelligence in Start-ups

Whether inventory management, financial accounting or storage costs, companies need access to up-to-date and reliable data at all times to plan well. In recent years, BI systems are increasingly being used by start-ups as well. Especially for data integration (ETL), data storage (Data Warehouse), data preparation or analysis (OLAP) and data presentation (planning, reporting). Reporting solutions eliminate the need for a manual and error-prone gathering of data from a variety of sources: all metrics are automatically merged into a single system and correlated in any desired combination. Thus, business intelligence has become an indispensable basis for decision-making in companies. Hardly any large enterprise today works without a BI system. Due to the implementation effort and the corresponding costs of the complex solutions, these seemed to be made only for large companies.

Especially for start-ups, BI systems offer a huge competitive advantage. For a company without a BI system, the development of a report requires enormous effort: tons of numbers are extracted manually from Excel spreadsheets, billing systems, and other programs to be saved in a new file. This goes through the hands of various employees. Such a procedure not only costs valuable working time, but it also carries the risk of severe transmission errors. If you instead use a systematic BI solution, the previously manually processed data is extracted and handled fully automatically. This significantly reduces both the risk of errors and saves human resources. Especially in start-ups, where individual employees often manage several areas of responsibility, this means a considerable relief.

The reaction times of the company can thus be shortened many times over. For as soon as a detailed search of company figures is no longer necessary, the time saved can be invested directly in the analysis. What are the current production and storage costs? Which product currently achieves the strongest sales? These questions can be answered with the help of a BI system without delay. Only those who know their numbers can react quickly and efficiently to the market. This ability to respond quickly is of particular importance to start-ups, as the planning horizon is usually much shorter than for a large company.

Even with negotiations with suppliers and customers, you are always one step ahead with BI: How high may the volume discount of a major customer be? Which supplier can achieve greater savings? Daily updated key figures provide the ideal basis for optimally positioning yourself in negotiations. Another central added value: A Business Intelligence solution is an effective early-warning system. Whether rising storage or falling production costs: If you always keep an eye on company figures, you will immediately notice changes. The analysis goes far beyond the actual daily numbers. Thanks to the comprehensive database, meaningful simulations are also available. Threatening sales or profit slumps are thus recognized at a very early stage and can be purposefully prevented or at least mitigated. For example, well-prepared and valid corporate numbers are also one indispensable basis for discussions with banks. Lending is often tied to regular reporting: with the help of a well-founded database, the desired information can not only be compiled quickly and easily. They can also be converted into meaningful formats, such as bringing graphics, dashboards or similar. Substantial forecasts create additional confidence.

Today, the spread of business intelligence in start-ups has increased significantly compared to previous years. Thus, in 2007, only half of the start-up companies had a BI application in use. Today, the market situation has changed fundamentally. Nearly 83 percent currently use a BI solution, with the remaining 17 percent using operational systems (such as ERP, CRM or SCM) and Excel. The investment in BI software is promoted not least by the weaknesses of the alternatives. In addition to ERP systems and spreadsheet software, there is a lack of functionality for preparing reports, distributing reports and analyzing, planning and coordinating various planners. Furthermore, data management functions are typically missing - above all possibilities for the integration of data from different sources as well as the use of central data memories that can be accessed by multiple users.

Business intelligence is used in start-ups mainly by controlling (89 percent), followed by management (70 percent) and sales (57 percent). As shown in the following figure.

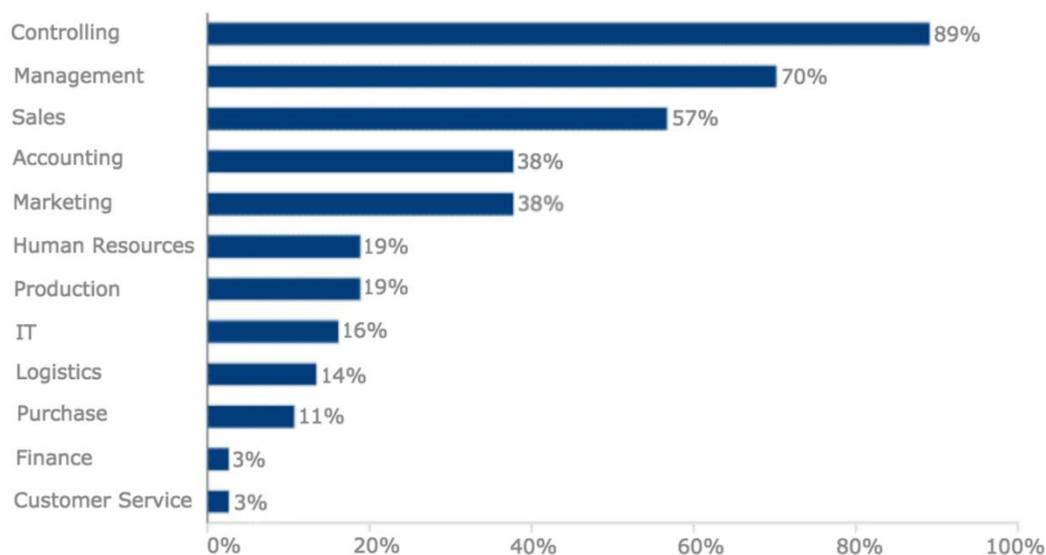

**Figure 2:** Application Areas of BI in Start-ups (BARC. (2012))

The areas of application of BI in start-ups show that they still tend to be used in classic departments. Controlling, as central processing, business data reporting, and planning entity uses BI software to support these tasks. Management uses BI data to extract decision-relevant information. Sales use BI primarily for customer-related analyzes.

Almost all start-up companies use business intelligence in data analytics (97 percent), 84 percent focus on report generation and distribution, 58 percent on planning and budgeting, and 49 percent on forecasting and rolling planning. However, more than half (60 percent) of the start-up companies plan to introduce improved management dashboards in the future. 46 percent want to invest more in forecasting and 39 percent in revised planning. The following figure shows the most common of the purpose of BI in start-ups.

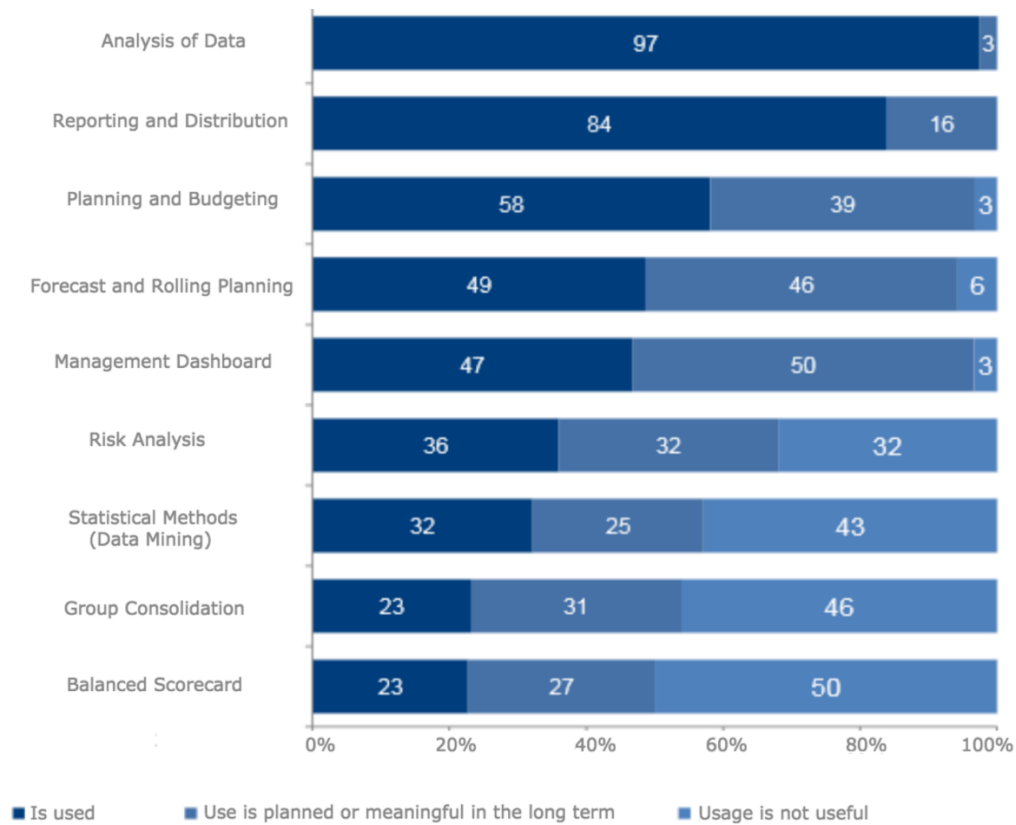

**Figure 3:** Usage purpose of BI in Start-ups (BARC. (2012))

It is hardly surprising that BI software is by far the most used for data analysis, reporting and distribution. Looking at future developments regarding planned use, planning and budgeting, forecasting and rolling planning can be named as primary fields of application. Users want to gain more information from existing data and process it further. The statistical methods (Data Mining), Group Consolidation and Balanced Scorecards do not appear to have found widespread adoption as a component of BI software in start-ups. Also, the information on planned or considered the beneficial use to the conclusion that they will not play an essential role in the future.

Start-ups also have significantly fewer people with BI know-how. About the introduction and use of BI, the corresponding skills must be available in the company. Here is a clear difference between medium-sized and larger companies, which would have to be resolved from the perspective of start-up companies to increase the BI deployment.

# 7. Success Factors for BI Projects in Start-ups

Successful BI projects are based on good content requirements profiles, which should be developed in a technical concept at the beginning of BI implementation. Due to the sometimes lacking know-how in the companies, it is useful here to draw on conceptual help from outside. There is little time for own staff aside from daily business for BI. Thus, the control and the success of a BI project are at risk. Well-functioning BI project coordination requires interdisciplinary competence between specialist knowledge (above all BI know-how) and leadership in the form of assertiveness, communication and coordination skills.

Key benefits of moving to BI solutions include better data integration, a combination of reporting and planning, and more significant flexibility and processing of large volumes of data.

A central success factor for a successful BI project is the involvement of the reporting and planning addressee in the preparation of the requirement profiles. A compact analysis of the current situation helps to get companies and their employees away from their point of view, to build on their strengths and thwart their weaknesses in the future. Another important success factor for a BI project is the specialist concept, which should contain all content, process, organizational and technical requirement profiles. Regarding contents, the value-adding and strategically relevant factors are to be identified. Also, content design and layout suggestions of the individual reports and the reporting structure and navigation in the reporting and planning system form important elements of the specialist concept. The technical content concept is the basis for the IT concept for the implementation of the BI solution. Above all, it was intended to determine the technical data sources of the upstream IT systems, to include a data supply concept and to describe the data modeling and query and report design technically.

The BI should not remain an isolated solution, then a data integration of all decision-relevant information from the variously available pre-systems is mandatory. The external support should be carried out according to the coaching principle and provide the necessary know-how transfer to BI for the project core team. As a result, the company remains independent of third parties in the medium term and can later develop the reporting system and plan within the company on its own. It often turns out that BI-based reporting and plan systems, which were high shaped by the employees themselves, are later better accepted and accepted by the company.

To secure the quality of the data, the introduction of suitable auxiliary instruments, e.g., To recommend reconciliation reports, account assignment and master data validations to be able to perform better plausibility checks and quality checks on the data. Also, software for verifying data quality is already being offered, using validation and transformation techniques such as parsing (syntax analysis). BI should not remain a tool for a select few, but the expansion of accessibility in BI-based reporting and planning should promote transparency, decision-making, and actionability throughout the company.

Reporting should go from compact core information to the top management level down to the decentralized decision-making areas in the enterprise and deliver decision-relevant financial and non-financial metrics to the controller. Data integration with BI makes the report generation and planning considerable shorten processes. Many manual processing steps in planning but also report preparation are eliminated by the data integration with BI, above all by the increased connection of the upstream systems.

Controlling as a central supporting body in the planning and reporting processes in the company uses more time for BI to prepare analyzes and to prepare prominent management comments and suggested measures for the management bodies. In addition, the admired Excel interface can remain as an integrative component in addition to new web-based application interfaces. Flexible analyzes help identify cause-and-effect relationships in the plant and its environment during live work in the system; this is supported by various output media and the many different access options to the information on portals, web access and mobile devices. The following table summarizes other benefits and disadvantages of using BI in start-ups.

**Table 1:** Benefits and Disadvantages of Using BI in Start-ups

| Benefits | Disadvantage |
|---|---|
| ❖ Functions for targeted and quick research that allow you to find and present relevant information in a current and consistent form<br>❖ Preparation of access to information according to factual and problem-related, possibly multidimensional criteria<br>❖ Illustration of tasks and problems using meaningful, realistic models<br>❖ Information processing and evaluation by powerful methods | ❖ Technology problems (problems of hardware technologies, development systems and BI software product itself)<br>❖ Development problems (engineering problems due to difficulties with the task area, with the tools and with the developers)<br>❖ Application problems (complexity of application areas)<br>❖ Maintenance problems (missing or deficient care and maintenance)<br>❖ Responsibility issues and acceptance issues (the user) |
| ❖ Presentation of results in understandable forms of presentation with multimedia techniques (data, texts, graphics, images, and language)<br>❖ Support of own work organization (for example scheduling)<br>❖ Promote cooperation in the joint resolution of tasks with collaborative systems<br>❖ Complementing operational application systems in the execution of process-oriented tasks with analytical functions<br>❖ Information and knowledge security in data and knowledge banks<br>❖ Help with the development of a functional knowledge management<br>❖ Integrate specific business applications into an existing information and analysis infrastructure<br>❖ Transparency and control of corporate processes<br>❖ Justification and traceability of management decisions made due to the permanent storage of information | ❖ Qualification problems (for all participants)<br>Problems of false expectation (wrong assessment of BI technologies) |

The start-ups have the task, before and during the use of BI, to recognize their benefits and disadvantages and to discuss them openly. The benefits that are offered must be identified and taken so that they lead to both operational and strategic vantages. The detriments, which often appear as weak signals, must be perceived and combated by containing or preventing adverse effects. Ultimately, the competence of start-up management can be measured by the extent to which it has been able to exploit the advantages and reduce the disadvantages or avoid the negative effects. Not only economic criteria, which are reflected in productivity, cost and output variables, can be used as benchmarks.

# 8. Results and Conclusions

The purpose of this paper was to give an insight into the current trend topic Business Intelligence in start-up companies. Business Intelligence is a relevant tool for successful management as well as start-ups. The BI systems convert the ever-increasing amounts of data of the individual operative systems into usable information. The BI tools for analysis and evaluation of the central database support the daily work of the executives and provide the basis for strategic decisions. Furthermore, the results of BI also serve the current company analysis. This can streamline business processes, improve customer and partner relationships, reduce costs, minimize risks, shorten processes, and gain competitive advantage.

Start-ups need to consider a wide variety of technical requirements and technical issues to successfully implement BI. Evaluations are becoming ever more extensive and contain a wide complexity of information about the company. A business model can be used to structure and efficiently develop a company-wide BI strategy. Also, developing a suitable BI strategy is the key to successful, long-term BI projects in start-up companies.

The market for business intelligence has evolved steadily in recent years and is also heavily dominated by the most important and well-known providers such as SAP, Oracle, IBM, SAS, Microsoft, and OpenSource. These companies offer software for start-ups that can successfully handle the five tasks of business intelligence, i.e., the collection, integration, storage, processing, and presentation of data. Also, the overall BI market is divided into the segments BI user tools (BI front ends) and BI data management (BI backend). BI front-ends include OLAP and multidimensional databases, data mining, dashboards, balanced scorecards, planning, reporting, and consolidation. The BI backend comprises software for data integration, data quality management, master data management, as well as relational or analytical databases in BI systems. In principle, many projects in the front-end area were initiated in start-ups, while investments in the back-end infrastructure are still reserved.

A rough market investigation shows that start-up companies like to buy open sources software vendors such as Jaspersoft, Pentaho, Jedox Palo, and SpagoBI. Because they can negotiate on an equal footing, have a local presence and support benefits, as well as know-how for specific tasks or industries. Another advantage of open source providers is that they offer a free version as well as a commercial version of their software. However, their commercial version can often be purchased cheaper than that of the mostly providers. The difference between the free version and the paid version is in most cases in advanced features and professional support.

The most frequently mentioned problems with BI use in start-ups are the lack of data quality and the lack of data integration from the previous systems. The lack of data integration shows that the start-up companies are not aiming for new isolated solutions with BI, but rather clear, company-wide BI solutions. The differences between start-ups and larger companies in the missing requirements from specialist departments are also obvious. There is, therefore, a need to catch up in start-ups in the preparation and the Matching Content Requirements Profiles for BI Solutions. Furthermore, to secure data quality in start-ups, the introduction of suitable auxiliary instruments, e.g., To recommend reconciliation reports, account assignment and master data validations to be able to carry out better plausibility checks and quality checks about the data. Also, software for verifying data quality is already being offered, using validation and transformation techniques such as parsing (syntax analysis).

It is anticipated that the adoption of BI solutions at start-up companies will continue to increase in the future, and thus more references from users will be available. As a result, BI applications will become more important and play an even more grave role in start-ups, as the demands continue to increase. The trend is towards ever faster and more detailed results, to be able to respond quickly to changes or to act better by prognoses in less time. Simple office applications, which are still used in many start-ups today, will no longer suffice as business intelligence solutions in the future. Especially very hard economic phases can survive companies with a professional Business Intelligence solution better than the companies, in which such a solution is not implemented. For start-ups, there are suitable professional solutions in the open source area.